\documentclass[preprint]{aastex631}

\shorttitle{Solar wind turbulence around Mars}
\shortauthors{Romanelli et al.}
\usepackage{amsmath}
\newcommand{\ua}{{\bf u}_{\text{A}}}
\newcommand{\uh}{{\bf u}}
\usepackage{ulem}

\begin{document}
\title{Variability of the Incompressible Energy Cascade Rate in Solar Wind Turbulence Around Mars}

\correspondingauthor{Norberto Romanelli}
\email{norberto.romanelli@nasa.gov}
\author[0000-0001-9210-0284]{Norberto Romanelli}
\affiliation{Department of Astronomy, University of Maryland, College Park, MD, USA}
\affiliation{Planetary Magnetospheres Laboratory, NASA Goddard Space Flight Center, Greenbelt, MD, USA}
\author[0000-0002-1272-2778]{Nahuel Andr\'es}
\affiliation{Departamento de F\'{\i}sica, UBA, Ciudad Universitaria, 1428, Buenos Aires, Argentina}
\affiliation{Instituto de Astronom\'ia y F\'{\i}sica del Espacio, CONICET-UBA, Ciudad Universitaria, 1428, Buenos Aires, Argentina}
\author[0000-0002-2778-4998]{Gina A.  DiBraccio}
\affiliation{Planetary Magnetospheres Laboratory, NASA Goddard Space Flight Center, Greenbelt, MD, USA}

\begin{abstract}
We present a statistical analysis on the variability of the incompressible energy cascade rate in the solar wind around Mars, making use of an exact relation for fully developed turbulence and more than five year of Mars Atmosphere and Volatile EvolutioN (MAVEN) observations. Using magnetic field and plasma data, we compute the energy cascade rate in the magnetohydrodynamics (MHD) scales in the pristine solar wind. From our statistical results we conclude that the incompressible energy cascade rate decreases as the Martian heliocentric distance increases, for each of the three explored Martian years. Moreover, we show that the presence of proton cyclotron waves, associated with the extended Martian hydrogen exosphere, do not have a significant effect in the nonlinear cascade of energy at the MHD scales.

\end{abstract}

\section{Introduction}

Turbulence is a unique multi-scale physical process present across the Universe, from the current of a river to the intergalactic medium \citep{Po2018,Al2018}. For fully developed turbulence, the plasma flow contains kinetic and magnetic fluctuations populating a wide range of spatial and temporal scales. In the so-called inertial range, decoupled from the injection large-scales and the dissipation small-scales, the total energy takes part in a cascade process across different scales. This particular process can be characterized by a scale-independent energy cascade rate $\varepsilon$, which represents the amount of energy per unit time across the inertial range. This $\varepsilon$ can be estimated using exact relations from fluid models, which involve correlation functions of the turbulent variables \citep[see,][]{Po2001}. For incompressible magnetohydrodynamics (MHD) turbulence, assuming statistical homogeneity and isotropy, \citet{P1998a,P1998b} have derived an expression for $\varepsilon$ as a function of two point spatial correlation functions of the velocity and magnetic fields. This theoretical result have been validated using numerical simulations \citep[e.g.,][]{Mi2009,W2010,A2018b}, in situ measurements \citep[e.g.,][]{SV2007,St2011,Co2012,H2017a,AN2021} and it has been extended to different plasma models, including compressible models, sub-ion scale effects and different thermodynamic closures \citep[e.g.,][]{Ga2008,B2013,A2017b,A2018,S2021a}. One such environment where the theoretical results are able to be tested with in-situ measurements is the case of solar wind turbulence around Mars, where ample data are now available.

 \citet{AN2020} computed the incompressible energy cascade rate upstream from Mars. More specifically, using two months of Mars Atmosphere and Volatile EvolutioN (MAVEN) data near the mission's first Martian perihelion and two months of data near the aphelion, the authors analyzed magnetic and solar wind plasma observations to conclude that the nonlinear cascade of energy is amplified when proton cyclotron waves (PCWs) are present in the plasma, near the Martian perihelion. In addition, using MAVEN high cadence magnetic observations from November 2014 to April 2016, \citet{Ru2017} characterized the magnetic energy spectra in several regions of the Martian magnetosphere and the surrounding solar wind. More specifically, the authors estimated the spectral indices at those regions, their variability with time and the potential connection with  Mars' seasonal variability and the PCWs occurrence rate.
 
In this context, proton cyclotron waves are ultra-low frequency electromagnetic plasma waves with observed frequencies (in the spacecraft frame) near the local proton cyclotron frequency and  whose most likely source are newborn planetary protons, a result from the ionization of the Martian extended hydrogen (H) exosphere \citep[e.g.,][]{R1990,B2002,M2004,R2013,Ru2016,Ro2016,R2018c,Romeo2021}. It is worth emphasizing that, although this term (proton cyclotron waves) has been used very often to make reference to them, they are not associated with the ion cyclotron wave mode. Temporal variability in the PCWs occurrence rate has been reported based on Mars Global Surveyor magnetic field observations \citep{Be2013,R2013} and more recently with MAVEN Magnetometer (MAG) measurements \citep{Con2015,J2015,Ro2016,Romeo2021}. By analyzing MAVEN  MAG observations between October 2014 and March 2016, \cite{Ro2016} reported that the PCW occurrence rate upstream of the Martian bow shock varies with time and takes higher values near the Martian perihelion. \cite{Romeo2021} confirmed such annual periodicity of PCWs occurrence rate upstream from the Martian bow shock for three Martian perihelia, analyzing MAVEN MAG and Solar Wind Ion Analyzer (SWIA) data between October 2014 and February 2020. This long-term trend has been associated with the seasonal variability of the Martian H exosphere, that extends well beyond the Martian bow shock, and provides the source of newborn planetary protons \citep[e.g.,][]{bhattacharyya2015,chaffin2014,clarke2017,yamauchi2015,rahmati2017,Ha2017b,Halekas2020_waves,Halekas2021}. Such variability is, in turn, partly associated to the relatively high eccentricity of Mars's orbit around the Sun.
To the best of our knowledge, there is not a study addressing and decoupling the influence that PCWs and the solar wind evolution with heliocentric distance
have on the energy cascade rate.

In this article we seek to determine the relative contribution that PCWs and the Martian heliocentric distance have on the incompressible energy cascade rate $\varepsilon$ in the MHD scales. This would allow us to determine the factor(s) responsible for the increase of $\varepsilon$ and its potential relation with Mars' seasonal variability \citep{Ru2017,AN2020}. To do this, we analyze more than five years of MAVEN MAG and SWIA observations when MAVEN was in the pristine solar wind. Such a large data set is required to differentiate effects due to PCW's activity and the Martian orbital eccentricity, as a result of the observed anti-correlation between PCW occurrence  and Mars' distance from the Sun.

\section{Incompressible MHD Turbulence}\label{sec:model}

The three-dimensional incompressible MHD equations are the momentum equation for the velocity field {\bf u} (in which the Lorentz force is included), the induction equation for the magnetic field {\bf B}, and the solenoid condition for both fields. These equations can be written as,
\begin{align}\label{1} 
	&\frac{\partial \textbf{u}}{\partial t} = -\uh\cdot\boldsymbol\nabla\uh  + \ua\cdot\boldsymbol\nabla\ua - \frac{1}{\rho_0}\boldsymbol\nabla(P+P_M) + \textbf{f}_k  + \textbf{d}_k , \\ 	\label{2} 
    &\frac{\partial \ua}{\partial t} = - \uh\cdot\boldsymbol\nabla\ua + \ua\cdot\boldsymbol\nabla\uh + \textbf{f}_m + \textbf{d}_m , \\  \label{3} 
    &\boldsymbol\nabla\cdot\uh = 0, \\ \label{4} 
    &\boldsymbol\nabla\cdot\ua= 0
 \end{align}
where we have defined the incompressible Alfv\'en velocity $\ua\equiv\textbf{B}/\sqrt{4\pi\rho_0}$ (where $\rho_0$ the mean mass density), and $P_M\equiv\rho_0 u_\text{A}^2/2$ and $P$ are the magnetic and plasma thermal pressure. Finally, \textbf{f}$_{k,m}$ are,  respectively, a mechanical and the curl of the electromotive large-scale forcings, and $\textbf{d}_{k,m}$ are, respectively, the small-scale kinetic and magnetic dissipation terms \citep{A2016b}. 

Using Eq.~\eqref{1}-\eqref{4} and following the usual assumptions for fully developed homogeneous turbulence (i.e., infinite kinetic and magnetic Reynolds numbers and a steady state with a balance between forcing and dissipation \citep[see, e.g.][]{F2021a}, an exact relation for incompressible MHD turbulence can be obtained as,

\begin{align}\label{exactlaw0}
	-4\varepsilon&= \rho_0\boldsymbol\nabla_\ell\cdot\langle (\delta\uh\cdot\delta\uh+\delta\ua\cdot\delta\ua)\delta\uh - (\delta\uh\cdot\delta\ua+\delta\ua\cdot\delta\uh)\delta\ua\rangle,
\end{align}

where $\varepsilon$ is the energy cascade rate per unit volume \citep[see,][]{P1998a,P1998b}. Fields are evaluated at position $\textbf{x}$ or $\textbf{x}'=\textbf{x}+\boldsymbol\ell$; in the latter case a prime is added to the field. The angular bracket $\langle\cdot\rangle$ denotes an ensemble average \citep{Po2001}, which is taken here as time average assuming ergodicity. Finally, we have introduced the usual increments definition, i.e., $\delta\alpha\equiv\alpha'-\alpha$. 
It is worth mentioning that we do not have access to multi-spacecraft measurements, and therefore, it is necessary to assume some sort of symmetry to integrate Eq.~\eqref{exactlaw0} and be able to compute the $\varepsilon$ \citep[see,][]{St2011}. Thus, assuming full isotropy and the Taylor hypothesis (i.e., $\ell\equiv\tau U_0$, where $U_0$ is the mean plasma flow speed and $\ell=|\boldsymbol\ell|$ is the longitudinal distance), Eq.~\eqref{exactlaw0} can be integrated and expressed as a function of time lags $\tau$. Therefore, the isotropic energy cascade rate can be computed as,

\begin{align}\label{law_iso}
	\varepsilon &= \rho_0\langle [(\delta\uh\cdot\delta\uh+\delta\ua\cdot\delta\ua)\delta{u}_\ell - (\delta\uh\cdot\delta\ua+\delta\ua\cdot\delta\uh)\delta{u}_{A\ell}]/(-4\tau U_0/3)\rangle.
\end{align}

where $u_\ell={\bf u}\cdot{\bf \hat U}_0$ and $u_{A\ell}=\ua\cdot{\bf \hat U}_0$. Therefore, $\varepsilon$ is fully defined by velocity and magnetic field time increments that we can estimate from MAVEN observations.

\section{MAVEN Observations and selection criteria}
\label{sec:selcriteria}

To investigate the variability of the incompressible energy cascade rate with the PCW activity and Martian heliocentric distance $r$, we analyze MAVEN MAG and SWIA data between 10 October 2014 and 31 December 2019. By analyzing these $\sim$ 5 years of observations, we are able to explore the three peaks of the PCWs occurrence rate reported in \cite{Romeo2021} and expand upon the work performed by \cite{AN2020}. MAVEN MAG measurements used in the present article have a 32 Hz sampling frequency and an accuracy of $\sim$ 0.25 nT \citep{Con2015}. SWIA measures ion flux sampling an energy range between 25 eV/q and 25 keV/q with a field of view of 360$^\circ{}$ × 90$^\circ{}$ \citep{Ha2015b}. Here, we have analyzed the onboard computed solar wind proton density and velocity moments, whose cadence is 4 s. It is worth mentioning that these moments assume that all ions are protons, which is a very good approximation for the pristine solar wind \citep{Halekas2017}.

First, we identify $\sim$ 34.1 min time intervals when MAVEN was upstream and magnetically disconnected from the Martian bow shock  \citep{Gru2018}. In particular, the size of each interval ensures having at least one correlation time of the turbulent fluctuations \citep{Ma2018}. In addition, by considering intervals of 65536 measurements (32 Hz MAG cadence) we are able to compute the Power Spectral Density (PSD) using a Fast Fourier method. More specifically, we compute the PSD of the magnetic field component perpendicular to the mean magnetic field to determine if PCWs are present or not in each interval under analysis, following the methodology reported in \cite{Ro2016} and \citet{Romeo2021}. Given the large size of the analyzed data sets, no overlap between neighboring intervals is considered. 

We follow a conservative approach to ensure all identified intervals are located upstream from the Martian bow shock. We do this by increasing the semi-latus rectum associated with the bow shock fit in 30$\% $\citep{Gru2018}, thus taking into account the variability in the shock's location. We also determine if MAVEN was connected or not to the bow shock during each analyzed time interval making use of the so-called Solar Foreshock Coordinates \citep{Greenstadt1986}. We consider an event is disconnect if the mean value of MAVEN's distance to the IMF tangent line (parallel to the solar wind flow) for each interval is negative. This allow us to avoid the nominal conditions of the Martian foreshock, a highly perturbed region with significant wave-particle activity \citep[e.g.,][]{Meziane2017}.

Finally, to compute the correlation functions involve in the Eq.~\eqref{law_iso}, for each interval we perform a linear interpolation of 32 Hz MAG data to the SWIA plasma on-board moments, density and velocity,  (0.25 Hz cadence). To have a reliable estimate of $\varepsilon$, both its sign and its absolute value, we focus our study on events in which the relative number density fluctuation and IMF cone angle are equal or smaller than $20\%$ \citep[see, e.g.,][]{H2017a,AN2020,AN2021}. The IMF cone angle is defined as the angle between the IMF and the solar wind velocity, and is also known to affect different regions of the  Martian  magnetosphere, for instance, the foreshock location \citep[see, e.g.,][]{Meziane2017,Romanelli2014,Romanelli2015,Dong2015,Dubinin2017,Romanelli2020}. We also require the variability of $\varepsilon$ with $\tau$ be relatively smooth, by considering events where the maximum of $\Delta|\varepsilon|$ for $\Delta\tau = 4s$ is smaller than $10^{-15}$ J m$^{-3}$ s$^{-1}$.

\section{Results}\label{sec:results}

\begin{figure}
\begin{center}
\includegraphics[width=1\textwidth]{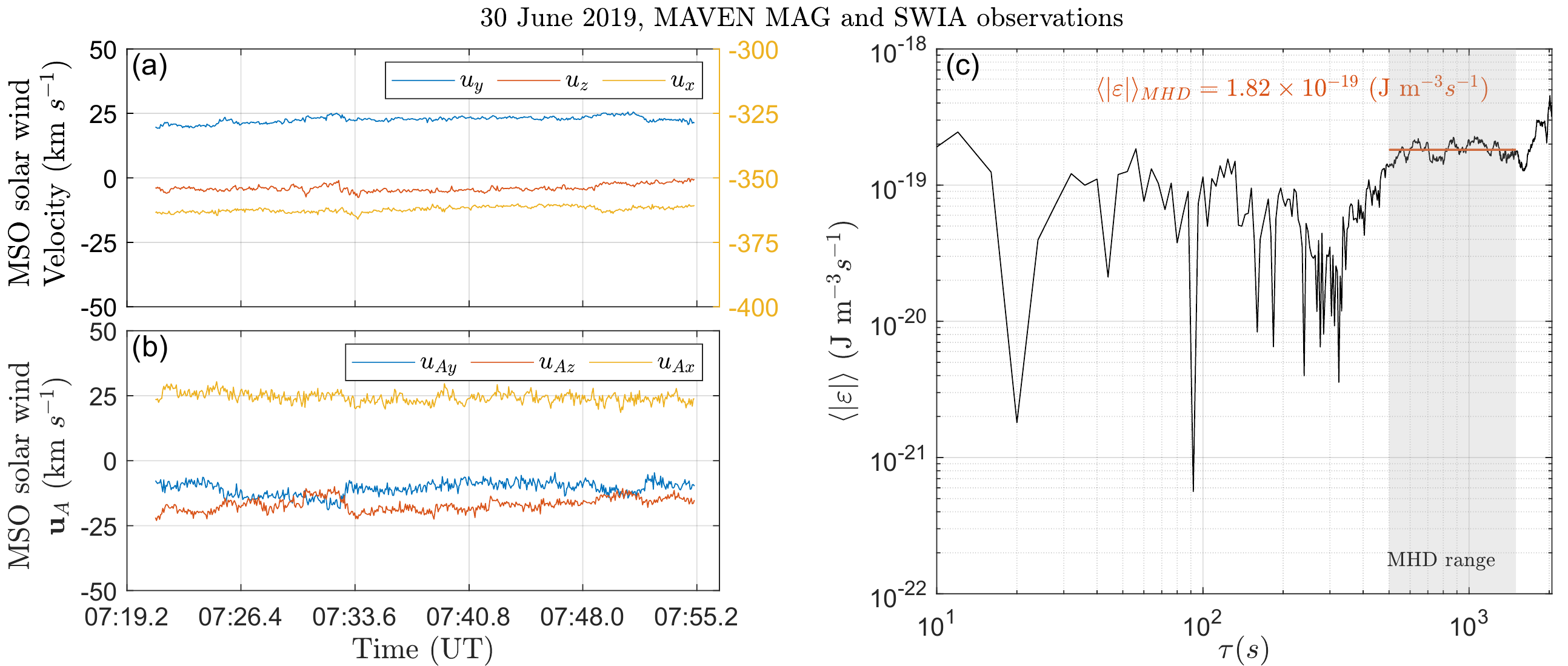}
\end{center}
\caption{Time series for an event observed in 30 June 2019 (a): Solar wind velocity components in the MSO coordinate system (left axis: $u_y, u_z$, right axis: $u_x$). (b): Solar wind Alfvén velocity components in the MSO coordinate system. (c) Energy cascade rate $\langle|\varepsilon|\rangle$ (absolute value) as a function of the time lag $\tau$.}
\label{fig1}
\end{figure}

Figure \ref{fig1} presents an example of an analyzed event, displaying MAVEN MAG and SWIA observations obtained on 30 June 2019, between 07:20:56 and 07:55:02 UT. Panels (a) and (b) show the solar wind and Alfv\'en velocity in the Mars Solar Orbital (MSO) coordinate system as a function of time, respectively. The  MSO coordinate system is centered at Mars and is defined as follows: the x-axis points towards the Sun, the z-axis is perpendicular to Mars’ orbital plane and is positive toward the ecliptic north, and the y-axis completes the right-handed system. As can be seen in panel (a), the solar wind velocity is mainly aligned with the x-axis (opposite sense) and displays a relatively constant value over the 31.4 min interval. Panel (b) shows the Alfv\'en velocity is also approximately constant, although it presents slightly more pronounced fluctuations. Figure \ref{fig1} (c) displays the absolute value of the incompressible energy cascade rate, $\langle|\varepsilon|\rangle$, as a function of $\tau$, for $\tau$ ranging between 4 s and 2044 s. We also determined the $\langle|\varepsilon|\rangle$-average in the MHD scale range. Following \cite{AN2020}, $\langle|\varepsilon|\rangle_{MHD}$ is computed as the average $\langle|\varepsilon|\rangle$ for $\tau$ ranging between 5$\times10^2$ s and 1.5$\times10^3$ s (shaded grey area). The $\langle|\varepsilon|\rangle_{MHD}$  value for this event is $1.8\times10^{-19}$ J m$^{-3}$ s$^{-1}$, two to three orders of magnitude smaller than the corresponding value observed in the solar wind at 1 au, $\langle|\varepsilon|\rangle_{MHD}\sim \times10^{-16}-\times10^{-17}$ J m$^{-3}$ s$^{-1}$ \citep{H2017a,AN2021}. The event shown in Figure \ref{fig1} took place when Mars' solar longitude ($L_s$) was equal to $46^\circ{}$, i.e., only $\sim$ 2 months before Mars reached aphelion on 26 August 2019 ($L_s=71^\circ{}$), partly explaining the relatively low value of  $\langle|\varepsilon|\rangle_{MHD}$. It is worth mentioning that we are only considering the magnitude of $\varepsilon$ rather than its signed value. Signed cascade rates are relevant to study the direction (i.e., direct vs. inverse) of the energy cascade, which is beyond the scope of this article.

To determine potential effects associated with PCWs activity and changes in the Martian heliocentric distance, we consider the three perihelia (PH) and aphelia (AH) periods that took place in the studied data sets, following the definition reported in \cite{Romeo2021}. Thus, the corresponding time periods for the PH1, PH2, and PH3 intervals are October 15,  2014–March 26, 2015, September 1,  2016–February 10, 2017, and July 20,  2018–December 29, 2018, respectively. Similarly, the time periods for the AH1, AH2, and AH3 intervals are August 31, 2015–April 9, 2016, July 18, 2017–February 25, 2018, and June 5, 2019–December 31, 2019, respectively. Hereafter, PH corresponds to the PH1, PH2 and PH3 intervals together, and AH corresponds to the AH1, AH2 and AH3 intervals.

\begin{figure}
\begin{center}
\includegraphics[width=0.95\textwidth]{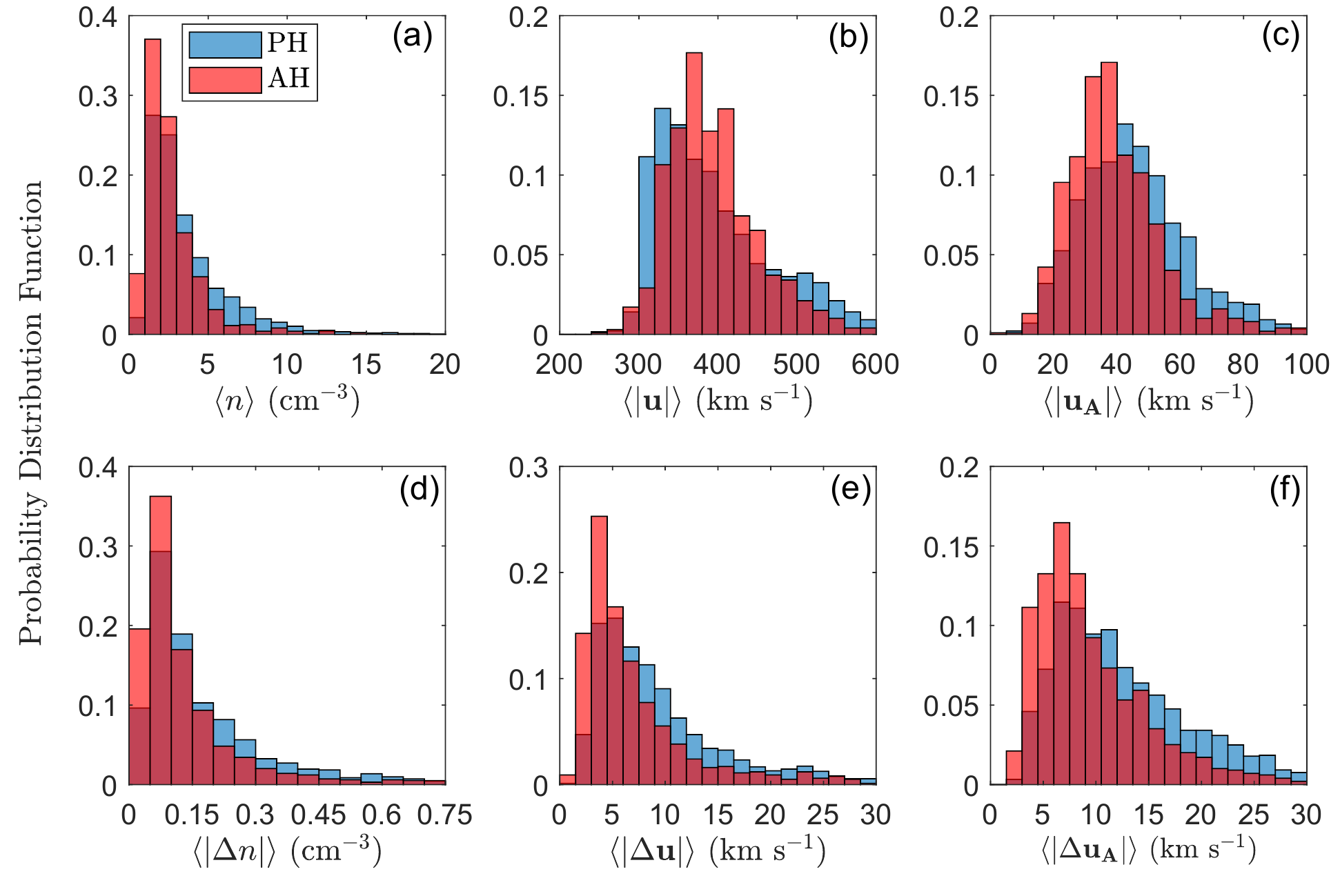}
\end{center}
\caption{(a)–(c) Normalized probability distribution function for the solar wind number density, velocity and Alfvén velocity absolute values, respectively. (d)–(f) Normalized probability distribution function for the corresponding fluctuations.}
\label{fig2}
\end{figure}

Figure \ref{fig2} shows the probability distribution function of the solar wind plasma density, velocity and Alfv\'en velocity absolute values, and their respective mean fluctuations for the PH and AH intervals, regardless of the PCW activity occurrence. The six distributions are similar, although PH conditions are characterized by slightly higher fluctuation values and by slightly denser solar wind plasma and higher Alfv\'en velocity absolute values. These differences are a result of the decreasing heliocentric distance and are in agreement with previous reports for Mars \citep{AN2020}. The analogous distributions for the perihelion interval, grouped depending on the PCWs activity (with and without PCWs and all) do not present significant differences (not shown). We do not consider the distribution for AH with PCWs, as there are not sufficient events. The reader is referred to \cite{Romeo2021} for a comprehensive assessment on this matter. 

Figure \ref{fig3}(a) shows the probability distribution function of log$\langle|\varepsilon|\rangle_{MHD}$ for events that took place in the PH interval with (in blue)  and without PCWs detection (in red). As can be seen, both distributions are highly similar ranging from $\sim-19$ to $\sim-15$, with $\langle|\varepsilon|\rangle_{MHD}$ median values equal to $1.4\times10^{-17}$ J m$^{-3}$ s$^{-1}$ and $1.5\times10^{-17}$ J m$^{-3}$ s$^{-1}$, respectively. Figure \ref{fig3}(b) shows the probability distribution function of log$\langle|\varepsilon|\rangle_{MHD}$ for events that took place in the PH interval (in blue)  and in the AH interval (in red), regardless of the PCWs activity. We find a clear shift between both distributions with larger values for PH conditions. The $\langle|\varepsilon|\rangle_{MHD}$ median value is equal to $1.5\times10^{-17}$ J m$^{-3}$ s$^{-1}$ and $4.6\times10^{-18}$ J m$^{-3}$ s$^{-1}$, respectively. 

Thanks to the large analyzed data set, we can also investigate the variability of $\langle|\varepsilon|\rangle_{MHD}$ 
with time and heliocentric distance. Figure \ref{fig4}(a) displays $\langle|\varepsilon|\rangle_{MHD}$ as a function of time, between 10 October 2014 and 31 December 2019, for time intervals that satisfied the conditions specified in Section \ref{sec:selcriteria}. Data gaps are mainly associated with time intervals where MAVEN did not sample the solar wind due to orbit precession \citep[see,][]{Romeo2021}. The right vertical axis corresponds to Mars heliocentric distance $r$, shown as a function of time in orange. As can be seen there is a appreciable amplification of $\langle|\varepsilon|\rangle_{MHD}$ around the  three explored Martian perihelia (vertical blue dashed lines).  Figure \ref{fig4}(b) displays $\langle|\varepsilon|\rangle_{MHD}$ as Mars heliocentric distance and therefore combines all data for the five explored years. This allow us to overcome some of the limitations due to the data gaps present in panel (a). We find that there is a clear anti-correlation between $\langle|\varepsilon|\rangle_{MHD}$ and $r$, with $\langle|\varepsilon|\rangle_{MHD}$ taking smaller values as Mars distance to the Sun increases.

\section{Discussion and Conclusions}\label{sec:conclusions}

In this work we investigate the variability of the incompressible energy cascade rate $\langle|\varepsilon|\rangle_{MHD}$ at the MHD scales in front the Martian bow shock. We analyze more than five years of MAVEN magnetic field and plasma observations in the pristine solar wind, that is, upstream and magnetically disconnected from the Martian bow shock. Thanks to this large statistical set, we explore the probability distribution function of $\langle|\varepsilon|\rangle_{MHD}$ for different solar wind conditions and three Martian perihelia and aphelia. Making use of four months of MAVEN MAG and SWIA data, \cite{AN2020} reported an amplification in $\langle|\varepsilon|\rangle_{MHD}$ when PCWs were detected in the pristine solar wind. From our large statistical study, we conclude that the nonlinear energy cascade rate at the MHD scales increases due to Mars's heliocentric distance decrease. Also, we find that PCWs do not have a significant effect on $\langle|\varepsilon|\rangle_{MHD}$. Therefore, we infer that previous results reported in \citet{AN2020} have the same dependence and that, due to the relatively small analyzed data set in that work, this feature could not be identified.
 
\cite{Ru2017} performed a characterization of turbulence inside and upstream of the Martian magnetosphere by determining, for the first time, the spectral indices for the magnetic field power spectra and studying their variability as a function of space, time and frequency range. In particular, the authors reported seasonal variability of the spectral indices, mostly in the region upstream of the Martian bow shock, which they associated with effects due to seasonal variability of the proton cyclotron waves occurrence rate \citep[e.g., ][]{Ro2016,Romeo2021}.  Our results suggest that, despite the wave energy associated with PCWs, the non linear energy cascade rate (and therefore, the magnetic spectral index) in the solar wind is not significantly affected at the MHD scales by the PCW's activity. It is worth emphasizing that, these waves, observed at the local proton cyclotron frequency are most likely fast magnetosonic waves, generated as a result of the ion-ion right hand resonant instability \citep[e.g.,][]{B1991,Gary1993,R2013}. Moreover, upstream from Mars, this instability is due to the interaction between newborn planetary protons derived from the extended Martian hydrogen exosphere and the incoming solar wind. Thus, by taking into account the Doppler shift associated with the relative speed between Mars and the solar wind, the wave frequency (in the plasma rest frame) is expected to be on the order of 0.1 solar wind proton gyrofrequency ($\tau \sim 200$ s), that is in the kinetic range \citep[see, e.g., Figure 8.2 in][]{Gary1993}. 
These considerations and our results suggest that, although the energy associated with PCWs does not significantly affect the nonlinear cascade rate, these plasma waves can play a key role in the energy transfer between scales in the kinetic regime. Additional analysis is needed to confirm this behavior, making use of higher cadence plasma observations and a theoretical model applicable to solar wind turbulence in these scales \citep{Ga2008,A2018}. Complementary studies should also be performed in different solar system planetary foreshocks. Indeed, these regions host many wave plasma modes that can potentially influence the non-linear energy transfer in various ways, as the solar wind conditions change with the distance to the Sun \citep[e.g.,][]{Eastwood2005,Burgess2012,AN2013,AndresNahuel2015,Meziane2017,Romanelli2021}. However, this investigation is beyond the scope of the present study. As reported in \cite{AN2020}, the discrepancy between our results and the observed effects in the spectral indices reported by \cite{Ru2017} could be due to several factors. In particular, the window size (512 s)  considered in  \cite{Ru2017}  may not be large enough compared to the correlation time of turbulent fluctuations to be able to compute reliable power spectral density fits and associated turbulent indices \citep{Ma2018}. Another consideration to take into account is that, in the present study, we compute the energy cascade rate when the IMF cone angle is relatively stable, to have a robust estimate of $\varepsilon$, while this condition is absent in \cite{Ru2017}. 

\begin{figure}
\begin{center}
\includegraphics[width=.99\textwidth]{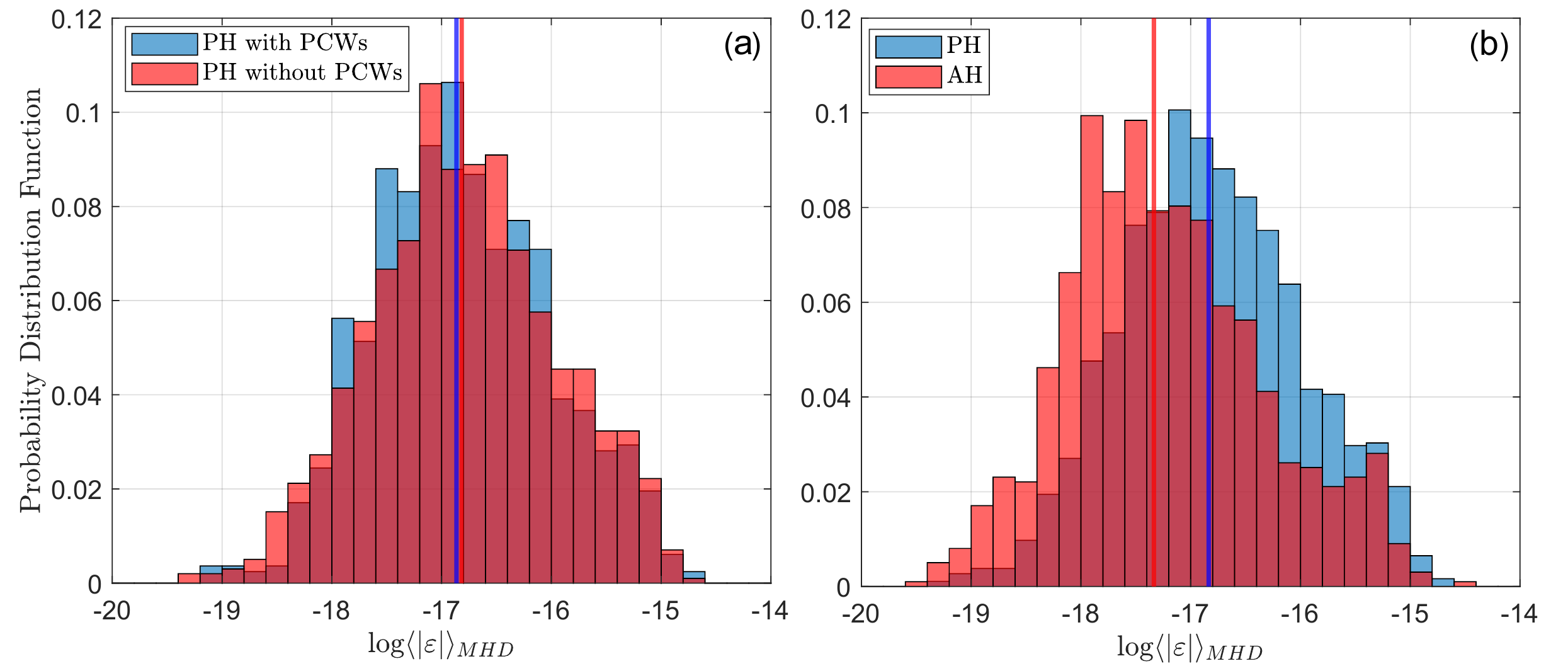}
\end{center}
\caption{Normalized probability distribution function of log$\langle|\varepsilon|\rangle_{MHD}$ for (a) Martian perihelion with PCWs (blue) and without PCWs (red), and for (b) Martian perihelion (with and without waves, blue) and aphelion conditions (with and without waves, red). In both panels, vertical solid lines (blue and red) correspond to the respective median of the distributions.
}
\label{fig3}
\end{figure}

\begin{figure}
\begin{center}
\includegraphics[width=1\textwidth]{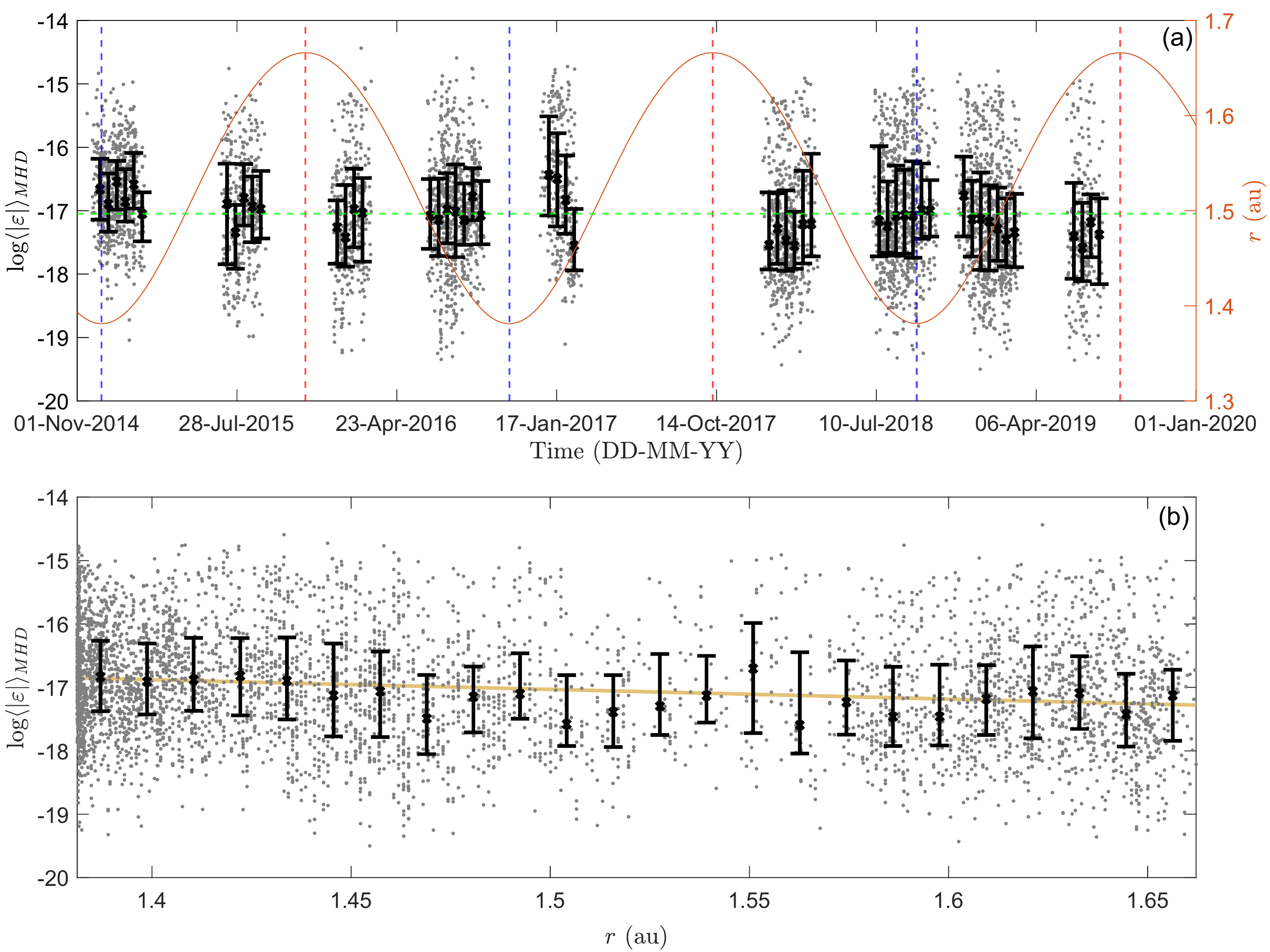}
\end{center}
\caption{(a, left vertical axis) Distribution of log$\langle|\varepsilon|\rangle_{MHD}$ as a function of time (gray dots), where black vertical bars correspond to the 25th, 50th, 75th percentiles for $\sim$ 14.3-day bins. Vertical blue and red dashed lines correspond to Martian perihelia and aphelia, respectively. Horizontal green dashed line corresponds to the median associated with all the analyzed events. (a, right vertical axis): Mars heliocentric distance $r$ as a function of time. (b) Distribution of log$\langle|\varepsilon|\rangle_{MHD}$ as a function of Mars heliocentric distance (gray dots), where black vertical bars correspond to the 25th, 50th, 75th percentiles for  $\sim$ 0.012-au bins. The yellow straight lines corresponds to the best linear fit.}
\label{fig4}
\end{figure}

We confirm that the nonlinear energy transfer rate upstream from the Martian bow shock is smaller compared to previous reports at smaller heliocentric distances \citep{H2017a,Ba2020, AN2021}. Indeed, we find the  median value of the  $\langle|\varepsilon|\rangle_{MHD}$-distribution varies between $1.5\times10^{-17}$ J m$^{-3}$ s$^{-1}$ and $4.6\times10^{-18}$ J m$^{-3}$ s$^{-1}$, between Martian perihelion and aphelion, respectively. Such values are at least an order of magnitude weaker when compared to reports based on observations of solar wind turbulence around Earth \citep{H2017a}. Also, Figure \ref{fig4} shows the median of the nonlinear cascade of energy  at MHD scales decreases with the Martian heliocentric distance in a relatively monotonous  manner. In addition, analysis of Parker Solar Probe measurements, have recently shown that the energy cascade rate is several orders of magnitude larger close to the Sun, when compared  their terrestrial counterparts, suggesting the presence of injection processes taking place near the Sun \citep{Ba2020,AN2021}. It is worth emphasizing that the absence of any significant effect of PCWs on $\langle|\varepsilon|\rangle_{MHD}$ suggest that the previous comparison are meaningful regardless of Mars orbital position around the Sun. Indeed, our results suggest that  solar wind turbulence around Mars (in the MHD scales) is not affected by PCW's presence nor by the newborn planetary protons (their most likely source) present upstream of the Martian bow shock. Finally, we report that the clear shift in the $\langle|\varepsilon|\rangle_{MHD}$ probability distribution functions for PH and AH intervals (shown in Figure \ref{fig3}(b)) is also observed for each pair of Martian perihelion and aphelion present in the analyzed time interval (not shown). In addition, we do not find a significant difference by the comparing the $\langle|\varepsilon|\rangle_{MHD}$ distributions for different perihelia and different aphelia, suggesting there is not a significant influence associated with the solar cycle. However, a larger data set is needed to properly analyze at least one complete solar cycle period.

\begin{acknowledgments}
The MAVEN project is supported by NASA through the Mars Exploration Program. N.R. is supported through a cooperative agreement with Center for Research and
Exploration in Space Sciences \& Technology II (CRESST II) between NASA Goddard Space Flight Center and University of Maryland College Park under award number
80GSFC21M0002. N.A. acknowledges financial support from the following grants: PICT 2018 1095 and UBACyT 20020190200035BA. MAVEN data are publicly available through the Planetary Data System (\url{https://pds-ppi.igpp.ucla.edu/index.jsp}).
\end{acknowledgments}


\bibliographystyle{aasjournal}

\end{document}